\documentclass[12pt]{iopart}

\usepackage{latexsym}

\def\cH{{\cal H}} 
\def\cR{{\cal R}}

\def\Tr{{\rm Tr}}
\def\d={\buildrel \rm def \over =}

\def\ket#1{\mid~\!\!\!{#1}~\!\!\rangle}
\def\bra#1{\langle~\!\!{#1}~\!\!\!\mid}

\begin{document}
\title[\bf Hermitian Schmidt Decomposition and Twins
of Mixed States]{\bf Hermitian Schmidt
Decomposition and Twin Observables of
Bipartite Mixed States}
\author{F Herbut\footnote[1]{E-mail:
fedorh@infosky.net}}
\address{Faculty of Physics, University of
Belgrade, POB 368, Belgrade 11001,
Yugoslavia and The Serbian Academy of
Sciences and Arts, Knez Mihajlova 35,
11000 Belgrade}

\date{\today}

\begin{abstract}
Study of mixed-state quantum
correlations  in terms of
opposite-subsystem observables the
measurement of one of which amounts to
the same as that of the other, of
so-called twins, is continued. Twin
events that imply biorthogonal mixing
of states, called "strong twin events",
are studied. It is shown that for each
mixed state there exists a Schmidt
(super state vector) decomposition in
terms of Hermitian operators, and that
it can be the continuation of the
mentioned biorthogonal mixing due to
strong twins. The case of weak twins
and nonhermitian Schmidt decomposition
is also investigated. For separable
states a necessary and sufficient
condition for the existence of
nontrivial twins is derived.
Utilization of the Hermitian Schmidt
decomposition for finding all twins is
illustrated in full detail in the case
of the two spin-one-half-particle
states with maximally disordered
subsystems (mixtures of Bell states).
It is shown that only rank two mixtures
have nontrivial twins.
\end{abstract}


\submitted \maketitle
\section{Introduction}
\rm Nowadays one distinguishes sharply
between separable bipartite mixtures,
which are quasiclassically correlated,
and nonseparable ones, endowed with
entanglement, a purely quantum
property. (A good example of the latter
is the case of correlated pure states.)
The term "quantum correlations" is used
in the generic sense comprising both
quasiclassical correlations and
entanglement.

It was claimed in a recent
investigation \cite{HD00} that the
study of quantum correlations through
twin observables, or shortly {\it
twins}, is expected to be important for
quantum communication and quantum
information theories because it is
believed to reveal some basic
properties of the correlations. Twin
observables are opposite-subsystem
observables such that the (subsystem)
measurement of one of them amounts {\it
ipso facto} to a measurement also of
the other. Equivalently put, the
subsystem measurement of a twin gives
rise, on account of the quantum
correlations, to an orthogonal
decomposition of the state of the
opposite subsystem.

In bipartite mixed states it is easier
to {\it relate twins to quantum
correlations} than to entanglement
(though the latter is more important).
A quantitative measure of the former
is, what is called, von Neumann's
mutual information $$
C(\rho_{12})\equiv S(\rho_{12}|\rho_1
\otimes \rho_2)=S(\rho_1)+S(\rho_2)-
S(\rho_{12}), $$ expressed in terms of
the so-called relative (or conditional)
entropy, and, alternatively, in terms
of the von Neumann entropies of the
reduced statistical operators $\rho_i,
\enskip i=1,2$ (subsystem states) and
the von Neumann entropy of the
statistical operator (bipartite state)
$\rho_{12}$ itself. I denote it by
"$C$" because it was thus designated
and called "logarithmic correlation" by
Lindblad \cite{Lind}. He also made use
of the classical discrete mutual
information $I(A,B|\rho_{12})$ of two
arbitrary opposite subsystem
observables  $A$ and $B$ (with purely
discrete spectra) which were assumed to
be simultaneously measured in a quantum
state $\rho_{12}$. Then, utilizing the
spectral forms and the ensuing
probabilities
$$A=\sum_ka_kP_1^{(k)},\quad k\not=
k'\enskip \Rightarrow \enskip a_k\not=
a_{k'},\qquad
B=\sum_lb_lQ_2^{(l)},\quad l\not=
l'\enskip \Rightarrow \enskip b_l\not=
b_{l'};$$ $$p(k,l)\equiv \Tr [\rho_{12}
(P_1^{(k)}\otimes Q_2^{(l)})], \quad
p_k\equiv \sum_lp(k,l),\quad p_l\equiv
\sum_kp(k,l);$$ one defines the mutual
information $$H(A:B|\rho_{12})\equiv
H(p(k,l)|p_kp_l)=H(p_k)+H(p_l)-H(p(k,l)),$$
where $H(p_k)$ e. g. is the so-called
Gibbs-Boltzmann-Shannon entropy
$H(p_k)\equiv -\sum_kp_klogp_k$, etc.
Finally, Lindblad defined
$$I(A,B|\rho_{12})\equiv
supH(A:B|\rho_{12}),$$ where the {\it
supremum} was taken over all possible
choices of the observables.

Lindblad showed that
$$I(A,B|\rho_{12})\leq C(\rho_{12}),$$
and that $$C(\rho_{12})>0\quad
\Rightarrow \quad I(A,B|\rho_{12})>0$$
are always valid.

Thus, in all correlated states, i. e.,
in states in which $C(\rho_{12})>0$, or
equivalently, $\rho_{12}\not=
\rho_1\otimes \rho_2$, one can
understand part of the quantum
correlations in terms of simultaneous
subsystem measurements and their
maximal mutual information.

Now, twins take up a very special
position among the subsystem
observables, because, if $A$ and $B$
are twins then $$I(A,B|\rho_{12})=
H(p_k),$$ since $H(p_k)=H(p_l)=
H(p(k,l))$ due to
$p(k,l)=p_k\delta_{l,f(k)}$, where
$f(k)$ is a fixed bijection of the
values of $k$ onto those of $l$. This
is the case of {\it perfect
correlations}, called "lossless and
noiseless information channel" in
information theory.

The investigation of {\it twins} began
with {\it pure states} \cite{FV76}
\cite{VF84} $\rho_{12}\equiv
\ket{\Phi}\bra{\Phi}$. Surprisingly, a
necessary and sufficient condition for
a subsystem observable $A$ to have a
nontrivial twin was found in terms of
properties of $\rho_1$ alone (local
properties, cf (2a)). The opposite
subsystem observable $B$ that is the
twin of $A$ was, naturally, expressed
in terms of global properties of
$\ket{\Phi}$. These were in a simple
way given in terms of an operator
(called correlation operator cf (11))
mapping the range of $\rho_1$ onto that
of $\rho_2$. It was defined by
$\ket{\Phi}$. This operator is most
practically handled in terms of the
so-called  {\it Schmidt decomposition}
\cite{Per} because it is precisely the
(antiunitary) operator determining
which characteristic vector of $\rho_2$
should appear in the same term as that
of $\rho_1$ in the mentioned
decomposition \cite{FV76} (cf (11)).

When twins were investigated in {\it
the mixed-state case} \cite{HD00}, the
mentioned condition (cf (2a)) was found
to be only necessary. Actually, a
sufficient condition for $A$ to have a
twin expressed as a property of
$\rho_1$ alone (a local property)
cannot exist because for every $\rho_1$
there is the uncorrelated state
$\rho_{12}\equiv \rho_1\otimes \rho_2$,
which does not have nontrivial twins.

Thus, global properties inherent in
$\rho_{12}$ have to be made use of at
the mentioned very first stage of
investigation of twins in the mixed
state case. It is not easy to "extract"
a minimal global property of
$\rho_{12}$ that does the job (as in
the pure state case).

It is a striking fact that the Schmidt
decomposition of state vectors can be
generalized to {\it all mixed states}.
It is the basic aim of this article to
investigate the relevance of this
decomposition to twins. It is proved
that the Schmidt decomposition of {\it
any bipartite mixed state}
 $\rho_{12}$ need not be expressed
in terms of some very general linear
operators, it can be given exclusively
in terms of Hermitian operators, which
can, in principle, be physically
interpreted as observables (cf Theorem
2 and Corollary 1).

The concept of strong twins, which are
closely connected with biorthogonal
decomposition of $\rho_{12}$ (cf
Theorem 1), is introduced as a step
towards the mentioned Hermitian Schmidt
decomposition of $\rho_{12}$. Also
nonhermitian Schmidt decomposition of
mixed states is studied (cf Theorem 3).

For mixtures of Bell states a Hermitian
Schmidt decomposition is given in the
literature (though not treated as
such). On this simple example the
problem of finding {\it all twins} is
easily solved  (cf Theorems 5 and 6) in
order to illustrate the relevance of
the Hermitian Schmidt decomposition to
extracting the sought for global
property inherent in $\rho_{12}$.

In the mentioned simple case it turns
out that rank four mixtures do not
allow nontrivial twins. This is not
surprising because it was shown in the
preceding study \cite{HD00} that
singularity of $\rho_{12}$ is a
necessary condition. But, surprisingly,
also rank three mixtures are shown to
have no nontrivial twins. This suggests
that perhaps a stronger necessary
condition, some kind of "sufficient
singularity", for the existence of
nontrivial twins could be found in the
general case. This will be followed up
elsewhere.

Relating twins to {\it separability} is
fully clarified in this study in terms
of a necessary and sufficient condition
for the existence of nontrivial twins
(cf Theorem 4). Relating twins to
 entanglement in the
mixed-state case, and to the
quantitative measures of entanglement
like the so-called entanglement of
creation and entanglement of
distillation \cite{Ben}, or the quantum
relative entropy \cite{Ved} and others
is an important open question that will
be hopefully treated in further work.

The study of twins having been pursued
in a number of mentioned articles is an
{\it ab ovo} approach, which is already
proved to be, in principle, relevant
and perhaps even important to quantum
information theory. It stands somewhat
apart from the mainstream
investigations. But it will be,
hopefully, connected up with the latter
as a result of further exploration.

\section{Preliminary Relations}
When a general, i. e., mixed or pure,
bipartite state (statistical operator)
$\rho_{12}$ is given, twins $(A_1,A_2)$
are algebraically defined as Hermitian
(opposite subsystem) operators
satisfying
$$
A_1\rho_{12}=A_2\rho_{12},\eqno{(1)}$$
where $A_1$ is actually $(A_1\otimes
I_2)$, $I_2$ being the identity
operator for the second subsystem, etc.
It was shown \cite{HD00} that (1)
implies
$$ [A_1,\rho_1]=0,\qquad
[A_2,\rho_2]=0\eqno{(2a,b)}$$
for the subsystem states (the reduced
statistical operators). (The symbols
$\Tr_i$, $i=1,2$, denote the partial
traces.) Relation (2a) is the mentioned
local necessary condition on $A_1$ to
have a twin.

If $P_1$ is a first-subsystem
projector, one can decompose the
statistical operator:
$$ \rho_{12}=P_1\rho_{12}+P_1^{\perp}
\rho_{12},\eqno{(3)}$$
where $P_1^{\perp}$ is the
orthocomplementary projector of $P_1$.
Let $(P_1,P_2)$ be a pair of nontrivial
twin events (twin projectors) for
$\rho_{12}$. In general, the terms on
the RHS are not even Hermitian. First,
we are going to investigate the more
important case when (3) is {\it a
mixture of states}.

\section{Strong twin projectors and
biorthogonal mixtures}
Let $(P_1,P_2)$ be a pair of nontrivial
twin projectors for a composite-system
statistical operator $\rho_{12}$.\\

\noindent {\it Remark 1.} Evidently,
either both terms on the RHS of (3) are
Hermitian or none of them is. They are
{\it Hermitian} if and only if the
projector $P_1$ (or equivalently,
$P_1^{\perp}$) {\it commutes} with
$\rho_{12}$:
$$ [P_i,\rho_{12}]=0,\qquad i=1,2,
\eqno{(4)}$$
(any one of the equalities implies the
other), as seen by adjoining the terms
in (3).

Hermiticity of the terms in (3) implies
that they are statistical operators (up
to normalization constants), i. e.,
that (3) is {\it a mixture}. Namely, if
(4) is valid, then idempotency leads to
$P_1\rho_{12}=P_1\rho_{12}P_1$, which
is evidently a positive operator. Since
$$ \Tr P_1\rho_{12}P_1\leq \Tr
\rho_{12}=1, $$
the operator has a finite trace.\\

\noindent {\it Definition 1.}
Nontrivial twin events (projectors) we
call either {\it strong twin events}
(projectors), if they satisfy (4), or
{\it weak} twin events (projectors) if
(4) is not satisfied.\\

A strong twin event $P_1$ implies a
mixture (3) of states that have a
strong property called biorthogonality.
To understand it, we first remind of
(ordinary) orthogonality of states.

If $\rho'$ and $\rho''$ are statistical
operators with $Q'$ and $Q''$ as their
respective range projectors, then one
has the known equivalences:
$$ \rho' \rho'' =0\quad \Leftrightarrow
\quad Q'Q''=0\quad \Leftrightarrow
\quad \cR(\rho')\perp
\cR(\rho''),\eqno{(5)}$$
where the last relation expresses
orthogonality of the ranges.

Any of the three relations in (5)
defines {\it orthogonality of
states}.\\

\noindent {\it Definition 2.} If
$$ \rho_{12}=w\rho_{12}'+
(1-w)\rho_{12}'',\qquad 0<w<1,
\eqno{(6)}$$
is a mixture of states such that
$$ \rho'_i\rho''_i=0,\qquad
i=1,2,\eqno{(7)}$$
where $\rho'_1\equiv \Tr_2\rho'_{12}$
etc. are the reduced statistical
operators, then we say that (6) is a
{\it biorthogonal mixture}.\\

To prove a close connection between
strong twin events and biorthogonal
mixtures, we need another known general
property of composite-system
statistical operators $\rho_{12}$:
$$ \rho_{12}=Q_1 \rho_{12}=
\rho_{12}Q_1=Q_2 \rho_{12}=
\rho_{12}Q_2,\eqno{(8)}$$
where $Q_i$ is the range projector of
the corresponding reduced statistical
operator $\rho_i$, $i=1,2$.\\

\noindent {\it Theorem 1.} If $P_1$ is
a nontrivial twin event, (3) is a {\it
biorthogonal mixture if and only if}
$P_1$ is a {\it strong twin event}.\\

\noindent {\it Proof. Sufficiency.} If
$P_1$ is a strong twin projector and
(6) is obtained by rewriting (3), then
$\rho'_{12}=P_1\rho'_{12}$ is valid,
and this implies $\rho'_1=P_1\rho'_1$
for the reduced statistical operator,
and, adjoining this, one arrives at
$\rho'_1=\rho'_1P_1$. On the other
hand, one has analogously
$\rho''_{12}=P_1^{\perp}\rho''_{12}$
implying
$\rho''_1=P_1^{\perp}\rho''_1$.
Finally,
$$ \rho'_1\rho''_1=(\rho'_1P_1)
(P_1^{\perp}\rho''_1)=0.$$

The symmetrical argument holds for the
second tensor factor.

{\it Necessity.} If (6) is a
biorthogonal mixture, then we define
$P_i\equiv Q'_i$, $i=1,2$, i. e., we
take the range projectors of the
reduced statistical operators of
$\rho'_{12}$ as candidates for our twin
projectors. On account of (8), we can
write (6) as follows:
$$  \rho_{12}=wQ_1'Q_2'
\rho_{12}'Q_1'Q_2'+ (1-w)Q_1''Q_2''
 \rho_{12}''Q_1''Q_2''.$$
Since in view of (5) biorthogonality
(7) implies $Q_i'Q_i''=0$, $i=1,2$, it
is now obvious that $P_1$ and $P_2$,
multiplying from the left $\rho_{12}$,
give one and the same operator, i. e.,
that they are twins, and it is also
obvious that they both give the same
irrespectively if they multiply $
\rho_{12}$ from the left or from the
right, i. e., that they are strong twin
projectors.

\hfill $\Box$

In view of (5), it is clear that
biorthogonal decomposition of a
statistical operator can be, in
principle, {\it continued}: If, e. g.,
$\rho'_{12}$ in a biorthogonal
decomposition (6) is, in its turn,
decomposed into biorthogonal
statistical operators and replaced in
(6), then any two of the new terms are
biorthogonal etc.

An extreme case of a biorthogonal
mixture is a {\it separable} one:
$$
\rho_{12}=\sum_kw_k\Big(\rho_1^{(k)}\otimes
\rho_2^{(k)}\Big),\eqno{(9)}$$
where
$$ \forall k:\quad w_k>0,\enskip
\rho_i^{(k)}>0,\enskip
\Tr\rho_i^{(k)}=1,\enskip i=1,2;\quad
\sum_kw_k=1$$
("$\rho >0$" denotes positivity of the
operator). This decomposition cannot,
of course, always  be carried out, but
examples are well known. For instance,
if one performs ideal nonselective
measurement of the z-component of spin
of the first particle in a singlet
two-particle state, one ends up with
$$ \rho_{12}\equiv
(1/2)\Big(\ket{z+}_1\bra{z+}_1 \otimes
\ket{z-}_2\bra{z-}_2\enskip +\enskip
\ket{z-}_1\bra{z-}_1 \otimes
\ket{z+}_2\bra{z+}_2\Big).$$
This is obviously a biorthogonal
separable mixture.

One wonders if, at the price of
relaxing the requirement of
statistical-operator terms as slightly
as possible, there could exist a {\it
general} decomposition into {\it
uncorrelated} terms (like in (9)).

To find an affirmative answer, we take
resort to the known case of general
(entangled or disentangled)
composite-system {\it state vectors}
and their Schmidt decompositions. Let
us sum up the sufficiently detailed
relevant information on this
\cite{FV76}.

{\it The Schmidt decomposition} of an
arbitrary pure state vector
$\ket{\Phi}_{12}$ of a composite system
is expressed in terms of {\it its
canonical entities}. They are the
following:

(i) {\it The reduced statistical
operators} (subsystem states) $\rho_1$
$\Big(\equiv
\Tr_2\ket{\Phi}_{12}\bra{\Phi}_{12}\Big)$
and $\rho_2$ (defined symmetrically)
are well known.

(ii) The spectral forms of the reduced
statistical operators are
$$ \rho_1=\sum_ir_i\ket{i}_1\bra{i}_1,
\quad
\rho_2=\sum_ir_i\ket{i}_2\bra{i}_2,
\quad \forall i:\enskip r_i>0.
\eqno{(10a,b)}$$
(Note that the positive spectra
-multiplicities included - are always
equal.)

(iii) Finally, the mentioned expansion
utilizes the (antiunitary ) {\it
correlation operator} $U_a$, which maps
the range $\cR(\rho_1)$ onto the range
$\cR(\rho_2)$. (Note that they are
always equally dimensional in the pure
state case). The correlation operator
is determined by $\ket{\Phi}_{12}$,
and, in turn, in conjunction with
$\rho_1$, it determines
$\ket{\Phi}_{12}$.

The {\it Schmidt decomposition} reads:
$$ \ket{\Phi}_{12}=\sum_ir_i^{1/2}
\ket{i}_1\otimes
\Big(U_a\ket{i}_1\Big)_2. \eqno{(11)}$$
The normalized characteristic vectors
$\ket{i}_2$ in (10b) may (and need not)
be chosen to be equal to
$\Big(U_a\ket{i}_1\Big)_2$.

In case of a state vector
$\ket{\Phi}_{12}$, the characteristic
relation (1) for twins reduces to
$$ A_1 \ket{\Phi}_{12}=A_2
\ket{\Phi}_{12}. \eqno{(12)}$$
 The corresponding twin $A_2$ then satisfies
$$ A_2=U_aA_1U_a^{-1}Q_2+
A_2Q_2^{\perp}, \eqno{(13)}$$
where $Q_2$ is the range projector of
$\rho_2$, and $Q_2^{\perp}$, its
orthocomplementary projector, projects
onto the null space of $\rho_2$.

One should note that, on account of the
commutation (2b), both the range and
the null space of $\rho_2$ are
invariant for $A_2$. Further, the
second term on the RHS of (13), or
rather the restriction of  $A_2$ to the
null space, which corresponds to it, is
completely arbitrary and immaterial for
the twin property (12), because it acts
as zero on $\ket{\Phi}_{12}$.
(Naturally, the symmetric claim holds
true for $A_1$ and $\rho_1$.)

\section{Hermitian Schmidt decomposition
of bipartite statistical operators}
It is well known that linear
Hilbert-Schmidt operators $A$ acting in
a Hilbert space, i. e., those with a
finite Hilbert-Schmidt norm $\Big(\Tr
A^{\dagger}A\Big)^{1/2}$, form a
Hilbert space in their turn. Writing
the operator $A$ as a (Hilbert-Schmidt)
supervector $\ket{A}$, the scalar
product is
$$ \bra{A}\ket{B}\equiv \Tr
A^{\dagger}B.$$

Since for every statistical operator
$\rho$, one has $\Tr\rho^2\leq 1$, it
is a  Hilbert-Schmidt operator.
Therefore, {\it every statistical
operator has a Schmidt decomposition}.

The trouble is that the operators that
take the place of the state-vector
tensor factors in the terms of (11),
which are the sought for
generalizations of the statistical
operators $\rho_i^{(k)},\enskip i=1,2$
in (9), are in general linear
operators. This might be a too wide
generalization. One wonders if one
could be confined to Hermitian
operators.

When we view the operators as
supervectors, then we must view {\it
adjoining} of operators as {\it an
antiunitary} operator the square of
which is the identity operator, i. e.,
which is an {\it involution}. Hence, we
denote adjoining by $V_1^{(a)}\otimes
V_2^{(a)}$ for a composite system.
Hermitian are the operators that are
{\it invariant} under the action of
this antiunitary involution.

Fortunately, the Schmidt decomposition
can always be expressed in terms of
Hermitian operators. We put this in a
more precise and a more detailed way.
But it is simpler to return to the
Hilbert space of state vectors to
perform some elaboration.\\

\noindent {\it Theorem 2.} Let
$V_1^{(a)}\otimes V_2^{(a)}$ be a given
antiunitary involution acting on
composite-system state vectors. One has
the equivalence:
$$\Big(V_1^{(a)}\otimes
V_2^{(a)}\Big)\ket{\Phi}_{12}=\ket{\Phi}_{12}
\quad \Leftrightarrow \quad
[\rho_i,V_i^{(a)}]=0,\enskip
i=1,2;\quad
V_2^{(a)}U_aV_1^{(a)}=U_a,\eqno{(14)}$$
where $\rho_i,\enskip U_a$ are the
above mentioned canonical entities of
$\ket{\Phi}_{12}$. (Note that in the
last relation we, actually, have the
restriction of $V_1^{(a)}$ to
$\cR(\rho_1)$.)\\

\noindent {\it Proof.} Let
$\ket{\Phi}_{12}$ be invariant under
the action of the antiunitary
involution. Then
$$ V_1^{(a)}\rho_1V_1^{(a)}=
V_1^{(a)}\Big(\Tr_2
\ket{\Phi}_{12}\bra{\Phi}_{12}\Big)
V_1^{(a)}=$$
$$ \Tr_2\Big(V_1^{(a)}
\ket{\Phi}_{12}\bra{\Phi}_{12}
V_1^{(a)}\Big)=\Tr_2\Big\{V_1^{(a)}
\Big[\Big(V_1^{(a)}\otimes
V_2^{(a)}\Big)
\ket{\Phi}_{12}\bra{\Phi}_{12}
\Big(V_1^{(a)}\otimes V_2^{(a)}\Big)
\Big] V_1^{(a)}\Big\}=$$
$$\Tr_2\Big( V_2^{(a)}
\ket{\Phi}_{12}\bra{\Phi}_{12}
V_2^{(a)}\Big)=\Tr_2
\ket{\Phi}_{12}\bra{\Phi}_{12}=
\rho_1,$$
and symmetrically for $\rho_2$. One has
to note that an antiunitary involution
equals its inverse and its adjoint.
Further, use has been made of some
known basic properties of partial
traces (which are analogous to the well
known ones for ordinary traces).

Commutation of $\rho_1$ with the
antiunitary involution $V_1^{(a)}$
allows one to choose the characteristic
basis $\{\ket{i}_1:\forall i\}$ of the
former spanning its range consisting of
vectors invariant under the action of
$V_1^{(a)}$ (cf \cite{Mes}).

Now, let us take the Schmidt
decomposition (11) in terms of such an
invariant basis. Then
$$ (V_1^{(a)}\otimes V_2^{(a)})
\ket{\Phi}_{12}=\sum_ir_i^{1/2}
\ket{i}_1\otimes
V_2^{(a)}\Big(U_a\ket{i}_1\Big)_2.$$
Since $\ket{\Phi}_{12}$ is assumed to
be invariant, it follows that also
$$ \ket{\Phi}_{12}=\sum_ir_i^{1/2}
\ket{i}_1\otimes V_2^{(a)}
\Big(U_a\ket{i}_1\Big)_2.$$
The second tensor factor in each term
is uniquely determined by the LHS and
the corresponding first tensor factor
(as a partial scalar product, cf
\cite{FV76}). Comparison with (11) then
shows that
$$ \forall i:\qquad
V_2^{(a)}U_a\ket{i}_1=U_a\ket{i}_1.$$
Since $\ket{i}_1=V_1^{(a)}\ket{i}_1$,
we further have
$$ V_2^{(a)}U_aV_1^{(a)}=U_a$$
as claimed.

Conversely, if the main canonical
entities are in the relation to the
antiunitary involutions as stated in
(14), then we can expand
$\ket{\Phi}_{12}$ in a characteristic
basis of $\rho_1$ spanning its range
that is invariant under the antilinear
operator. Then (11) immediately reveals
that, as a consequence,
$\ket{\Phi}_{12}$ is invariant under
$V_1^{(a)}\otimes V_2^{(a)}$. \hfill
$\Box$\\

\noindent {\it Corollary 1.} Every
composite-system statistical operator
$\rho_{12}$ has, after normalization, a
{\it Hermitian} Schmidt
decomposition.\\

\noindent {\it Proof.} Since every
$\rho_{12}$, being Hermitian, is
invariant under the antiunitary
involution $V_1^{(a)}\otimes
V_2^{(a)}$, Theorem 2 immediately
implies that $\rho_{12}$, upon super
vector normalization, has a Schmidt
decomposition in terms of Hermitian
operators.\hfill $\Box$\\

 Returning to a
biorthogonal mixture, one wonders if
one can {\it continue} such a
decomposition by writing each term in a
Hermitian Schmidt decomposition in
order to obtain the latter
decomposition for the entire
statistical operator. The answer is
affirmative on account of the
following:

Going back to (5), we can add a fourth
equivalent property.\\

{\it Proposition 1.} Two statistical
operators $\rho'$ and $\rho''$ are {\it
orthogonal if and only if they are
orthogonal as Hilbert-Schmidt
supervectors}.\\

{\it Proof.} It is obvious that
orthogonality (in the sense of (5))
implies Hilbert-Schmidt orthogonality.
To see the converse implication, we
make use of the fact that every
statistical operator has a purely
discrete spectrum \cite{Sim}, and we
decompose the statistical operators in
terms of characteristic vectors
corresponding to positive
characteristic values:
$$ \bra{\rho'}\ket{\rho''}=\Tr\rho'
\rho''=\Tr\sum_kr_k\ket{k}\bra{k}
\sum_j\bar r_j\ket{j}\bra{j}=$$
$$ \sum_k\sum_jr_k\bar r_j
|\bra{j}\ket{k}|^2.$$
Hence,
$$ \bra{\rho'}\ket{\rho''}=0\quad
\Rightarrow \quad \rho'\rho''=0$$
(cf the third relation in (5)).\hfill
$\Box$\\

 If $(A_1,A_2)$ is a pair of {\it twin
observables}, then the detectable parts
$A_i',\enskip i=1,2$, have a common
purely discrete spectrum $\{a_n:\forall
n\}$ (with, in general, different
multiplicities), and the corresponding
(detectable) characteristic projectors
$\{P_i^{(n)}:i=1,2\enskip \forall n\},$
are also pairs of twins \cite{HD00}.\\
\noindent {\it Definition 3.} If {\it
all} mentioned characteristic projector
pairs $(P_1^{(n)},P_2^{(n)})$ are
strong twin projectors, then
$(A_1,A_2)$ is a pair of {\it strong
twin observables}. If some of the
detectable characteristic twin
projectors are strong and some weak, we
say that we have {\it partially strong}
(or, synonymously, partially weak) twin
observables. If all the mentioned twin
projectors are weak, then we have a
{\it weak} pair of twin observables.\\

A pair $(A_1,A_2)$ of nontrivial twin
observables for $\rho_{12}$ is a pair
of strong ones if and only if
$$ [A_i,\rho_{12}]=0,\quad
i=1,2\eqno{(15)}$$
is valid. This is so because
commutation with all characteristic
projectors is equivalent to commutation
with the Hermitian operator itself,
and, if $P_1$ e. g. is a nondetectable
characteristic projector of $A_1$, then
one has commutation because
$$P_1\rho_{12}=(P_1Q_1^{\perp})\rho_{12}=0
=\rho_{12}(Q_1^{\perp}P_1)=\rho_{12}P_1$$
on account of (8).

Strong twin observables, by means of
their strong characteristic twin
projectors, lead to a generalization of
(3):
$$ \rho_{12}=\sum_nP_1^{(n)}\rho_{12}=
\sum_nw_n\rho_{12}^{(n)},\eqno{(16a)}$$
where
$$ \forall n:\qquad w_n\equiv \Tr
\rho_{12}P_1^{(n)},\quad
\rho_{12}^{(n)}\equiv (w_n)^{-1}
P_1^{(n)}\rho_{12}.\eqno{(16b)}$$
Naturally, if $P_1^{(n)}\rho_{12}=0$,
then $\rho_{12}^{(n)}$ is not defined.
Any two terms in (16a) are
biorthogonal.

Note that we utilize the entire
characteristic projectors, which are
the orthogonal sums of the detectable
and the nondetectable parts:
$P_1^{(n)}= (P_1')^{(n)}\oplus
(P_1'')^{(n)}$ parallelling
$\cH_1=\cR(\rho_1)\oplus
\cR^{\perp}(\rho_1)$ because
$(P_1')^{(n)}\rho_{12}=P_1^{(n)}\rho_{12}$.
\\

\noindent {\it Proposition 2.} If
$$ \rho_1^{(n)}\equiv
\Tr_2\rho_{12}^{(n)},$$
and symmetrically for $\rho_2^{(n)}$,
are the reduced statistical operators
of the terms in a biorthogonal mixture
(16a), then
$$
P_i^{(n)}\rho_i^{(n)}=\rho_i^{(n)},\quad
i=1,2,\eqno{(17a)}$$
or equivalently,
$$ \cR(\rho_i^{(n)})\subseteq
\cR(P_i^{(n)}),\quad
i=1,2.\eqno{(17b)}$$

{\it Proof.} On account of the
definition of (16a), one has
$P_i^{(n)}\rho_{12}^{(n)}=\rho_{12}
^{(n)}$. Taking the opposite-subsystem
partial trace, one obtains
$P_i^{(n)}\rho_i^{(n)}=\rho_i
^{(n)}\enskip i=1,2$.\hfill $\Box$\\

\noindent {\it Corollary 2.} If the
detectable part $A'_1$ of a twin
observable $A_1$ has a {\it
nondegenerate} characteristic value
$a_n$ corresponding to a strong
characteristic twin projector
$(P_1')^{(n)}=\ket{\psi^{(n)}}_1
\bra{\psi^{(n)}}_1,\quad
\ket{\psi^{(n)}}_1\in \cR(\rho_1)$,
then the term in the biorthogonal
mixture (16a) that corresponds to it
has the form
$$
w_n\ket{\psi^{(n)}}_1\bra{\psi^{(n)}}_1
\otimes \rho_2^{(n)},\eqno{(18)}$$
where $\rho_2^{(n)}$ is a
(second-subsystem) statistical operator
and (18) is a term in a final Hermitian
Schmidt decomposition of $\rho_{12}$.\\

Any biorthogonal decomposition of a
composite-system statistical operator
$\rho_{12}$ (into two or more terms)
can be continued in each term
separately into a Schmidt decomposition
of $\rho_{12}$ in terms of Hermitian
operators.

The biorthogonal decomposition is {\it
an intermediate step}. This is similar
to the case when we can partially
diagonalize the Hamiltonian of a
quantum system (due to some symmetry e.
g.). The diagonalization is then
continued separately with each
submatrix on the diagonal of the
Hamiltonian.

The continuation from a biorthogonal
mixture to a Hermitian Schmidt
decomposition can always be performed,
in principle, "by brute force":
diagonalizing the reduced statistical
superoperator $\hat \rho_1$ of the
normalized supervector
$\ket{\rho_{12}}$ (analogously as it is
done for an ordinary state vector), and
by finding an invariant basis for
$V_1^{(a)}$ in each characteristic
subspace thus obtained \cite{Mes}.

\section{Weak twins and nonhermitian
 Schmidt decomposition}

For the sake of completeness it is
desirable to investigate decomposition
(3) also for a weak nontrivial twin
projector $P_1$. First, we take an
analytical view of Theorem 1 to realize
that the biorthogonality of the two
terms in (3) is connected with the twin
property (strong or weak), and the
strong twin property corresponds to the
hermiticity of the terms. Let us put
this more precisely.\\

\noindent {\it Remark 2.} A
decomposition
$$ \rho_{12}=A_{12}+B_{12}$$
of a composite-system statistical
operator $\rho_{12}$ into two linear
operators is {\it biorthogonal} if
there exist two opposite-subsystem
projectors $(P_1,P_2)$ such that
$$ A_{12}=P_1A_{12}=P_2A_{12},\quad
0=P_1B_{12}=P_2B_{12};$$
$$
0=P_1^{\perp}A_{12}=P_2^{\perp}A_{12},\quad
B_{12}=P_1^{\perp}B_{12}=P_2^{\perp}B_{12}.
$$

It is clear from Theorem 1 that any
birthogonal mixture (of states) (6)
satisfies the condition given in Remark
2. Having in mind (3), it is also
evident that biorthogonality is
equivalent to the existence of a pair
of twin projectors (weak or strong).
Finally, the strongness property of the
twins is equivalent to the hermiticity
of the terms in (3), which results in
having statistical operator terms (and
a mixture).\\

\noindent {\it Theorem 3.} If
$(P_1,P_2)$ is a pair of {\it weak twin
projectors} for a composite-system
statistical operator $\rho_{12}$, then
the terms in (3) are super vectors, and
replacing each by a (nonhermitian)
 Schmidt decomposition, one obtains
 a decomposition of the same kind for the
 entire statistical operator.\\

\noindent {\it Proof.} Since in
$$ 1\geq \Tr \rho_{12}^2=\Tr \rho_{12}
P_1\rho_{12}\enskip +\enskip \Tr
\rho_{12} P_1^{\perp}\rho_{12}$$
the terms are nonnegative (as traces of
positive operators), the terms in (3)
are Hilbert-Schmidt operators, i. e.,
super vectors. Suppose we have
decomposed the first term in (3) in the
Schmidt way:
$$
P_1\rho_{12}=c\sum_ir_i^{1/2}A_1^{(i)}
\otimes B_2^{(i)},$$
where $c$ is a normalization constant
(because the statistical operator is
not a super state vector unless it is a
pure state). Since the LHS is invariant
under $P_1$, so is each first-subsystem
linear operator $A_1^{(i)}$, because
the second factors in the expansion
have unique corresponding first
factors. If we decompose also the
second term in (3) in the Schmidt way
$$P_1^{\perp}\rho_{12}=c'\sum_jr_j'^{1/2}
C_1^{(j)} \otimes D_2^{(j)},$$
then, analogously, invariance of each
factor $C_1^{(j)}$ under $P_1^{\perp}$
follows. This results in super vector
orthogonality:
$$ \forall i,j:\qquad \Tr
\Big[(A_1^{(i)})^{\dagger}C_1^{(j)}
\Big]= \Tr \Big[
(A_1^{(i)})^{\dagger}P_1P_1^{\perp}
C_1^{(j)}\Big]=0.$$
The symmetrical argument goes for the
second factors and $P_2$. Thus,
replacing both terms in (3) by their
nonhermitian  Schmidt decompositions,
we have biorthogonality between any
term of the first decomposition and any
term of the second one. Therefore, we
have a decomposition of the same kind
of the entire $\rho_{12}$.\hfill
$\Box$\\

It is now clear that also in the case
of weak twin projectors the
decomposition (3) can be {\it
continued}, but this time {\it to a
nonhermitian Schmidt decomposition}.

A nonhermitian Schmidt decomposition
need not be wild and far fetched from
the physical point of view. Let me
illustrate this by the obvious fact
that a Schmidt decomposition of a state
vector $\ket{\Phi}_{12}$
$$ \ket{\Phi}_{12}=\sum_ir_i^{1/2}
\ket{i}_1\ket{i}_2,\qquad \forall
i\not= i':\quad \bra{i}_p
\ket{i'}_p=0,\enskip p=1,2$$
immediately results in a nonhermitian
 Schmidt decomposition of the statistical operator
$\ket{\Phi}_{12}\bra{\Phi}_{12}$:
$$ \ket{\Phi}_{12}\bra{\Phi}_{12}=
\sum_i\sum_{i'}r_i^{1/2}r_{i'}^{1/2}
\ket{i}_1\bra{i'}_1\otimes
\ket{i}_2\bra{i'}_2.$$

 Finally, let us return to separable
mixtures.

\section{Nontrivial twin projectors for
separable mixtures}
Let (9) be a general separable mixture.
Let us clarify under what conditions it
has nontrivial twin events.\\

\noindent {\it Theorem 4.} A general
separable mixture (9) has a nontrivial
twin projector $P_1$ {\it if and only
if} the set of all values of the index
"$k$" is the union of two
nonoverlapping subsets, say, consisting
of "$k'$" values and of "$k''$" values
respectively, and, when (9) is
rewritten accordingly:
$$
\rho_{12}=\sum_{k'}w_{k'}\rho_1^{(k')}
\otimes \rho_2^{(k')}+\sum_{k''}w_{k''}
\rho_1^{(k'')}\otimes \rho_2^{(k'')},
\eqno{(19a)} $$
then one has biorthogonality between
the two groups of terms:
$$ \forall k',\enskip \forall
k'':\qquad
\rho_i^{(k')}\rho_i^{(k'')}=0,\quad
i=1,2.\eqno{(19b)}$$
Before we prove the theorem, we first
prove subsidiary results.\\

\noindent {\it Lemma 1.} Let
$$
\rho_{12}=\sum_mw_m\ket{\Psi^{(m)}}_{12}
\bra{\Psi^{(m)}}_{12}$$
be an arbitrary pure-state mixture.
Then, a pair of  subsystem observables
$(A_1,A_2)$ are twins for $\rho_{12}$
{\it if and only if} they are twins for
{\it all} pure-state terms.\\

\noindent {\it Proof. Necessity}
follows from the general result that
all twins of $\rho_{12}$ are also twins
of all state vectors from the
topological closure $\bar
\cR(\rho_{12})$ of the range of
$\rho_{12}$ (cf section 3, C1 in
\cite{HD00}). As well known, the
vectors
$\{\ket{\Psi^{(m)}}_{12}:\forall m\}$
span the mentioned subspace.

{\it Sufficiency} is obvious.\hfill
$\Box$\\

\noindent {\it Lemma 2.} Let
$$ \rho_{12}=\sum_kw_k\rho_{12}^{(k)}$$
be an arbitrary mixture. The pair
$(A_1,A_2)$ are twin observables for
$\rho_{12}$ {\it if and only if} they
are twin observables for {\it all} term
states $\rho_{12}^{(k)}$.\\

\noindent {\it Proof} is immediately
obtained from Lemma 1 if one rewrites
each term state as a pure-state
mixture.\hfill $\Box$\\

\noindent {\it Lemma 3.} An {\it
uncorrelated state} $\rho_1\otimes
\rho_2$ has only trivial twins.\\

\noindent {\it Proof} is an immediate
consequence of the fact that the tensor
factors of a nonzero uncorrelated
vector, say $a\otimes b$, are unique up
to an arbitrary nonzero complex number
$\alpha$, but if $a$ is replaced by
$\alpha a$, $b$ must be replaced by
$(1/\alpha )b$.

If two observables are twins for an
uncorrelated state, then
$$ A_1\rho_1\otimes
\rho_2=\rho_1\otimes A_2\rho_2.$$
If $A_1\rho_1=\alpha\rho_1$, then,
applying the above remark to
supervectors, one has
$\rho_2=(1/\alpha)A_2\rho_2$.\hfill
$\Box$\\

\noindent {\it Proof of Theorem 3} now
immediately follows from Lemma 2 and
Lemma 3. Namely, the two groups of
terms stated in the Theorem, make up
the two terms in (3).\hfill $\Box$\\

\noindent {\it Corollary 3.} Nontrivial
twin events of a separable mixture (9)
are {\it necessarily strong twin
events}.\\

\noindent {\it Corollary 4.} If
$(A_1,A_2)$ are nontrivial twin
observables for a separable mixture
(9), they are strong twin observables
(cf Definition 3), and the mixture
terms can be grouped into as many
biorthogonal groups of terms as there
are distinct characteristic values of
$A_1$ in $\cR(\rho_1)$ (generalization
of (19a,b)).\\

It is known that if a statistical
operator and a Hermitian operator
commute, then the corresponding state
can be written as a mixture so that
each term-state has a definite value of
the corresponding observable
\cite{Hro}. But, for the same
statistical operator, there are also
mixtures violating this.

To take an example, let us think of an
unpolarized mixture of spin-one-half
states: $\rho =(1/2)I$ (in the
two-dimensional spin factor space).
This statistical operator commutes with
$s_z$, nevertheless one can write down
the mixture
$$ \rho =(1/2)\Big(\ket{x,+}\bra{x,+}+
\ket{x,-}\bra{x,-}\Big)=(1/2)I,$$
in which the term-states do not have a
definite value of the z-component.

It is interesting that in the case of a
separable mixture with a nontrivial
twin observable it is necessarily its
term-states that have the sharp
detectable values of the corresponding
observable.\\

\section{States with Maximally Disordered
Subsystems}

 Now we turn to the example that is,
 for illustrative purposes, investigated
in this study, i. e., to states
(statistical operators) $\rho$ in
$C^2\otimes C^2$. We say that $\rho$ is
an MDS  state (one with maximally
disordered subsystems or rather
subsystem states) if $\rho_1=(1/2)I_1$
and $\rho_2=(1/2)I_2$. R. and M.
Horodecki have shown \cite{Hor} that
for every MDS state there exist unitary
subsystem operators $U_1$ and $U_2$
such that
$$ \Big(U_1\otimes U_2\Big)\rho
\Big(U_1^{\dagger}\otimes U_2^{\dagger}
\Big)=(1/4)\Big(I\otimes I+\sum_{i=1}^3
t_i \sigma_i\otimes \sigma_i\Big)\equiv
T, \eqno{(20)}$$
where $\sigma_i$, $i=1,2,3$, are the
well known Pauli matrices $\sigma_x,
\sigma_y$ and $\sigma_z$; and it is
seen from their place in the expression
if they are meant for the first or for
the second spin-one-half particle.

Further, they have shown that the
operator $T$ is a statistical operator
(a quantum state) if and only if the
vector $\vec t$ from {\bf R}$_3$ the
components of which appear in (20) is
not outside the {\it tetrahedron}
determined as the set of all mixtures
of the four pure {\it Bell states}:
$$ \ket{\psi^1_2}\equiv
(1/2)^{1/2}\Big(\ket{+}\ket{+}\enskip
{\buildrel - \over +}\enskip \ket{-}
\ket{-}\Big), \quad
\ket{\psi^3_0}\equiv
(1/2)^{1/2}\Big(\ket{+} \ket{-}\enskip
{\buildrel + \over -}\enskip \ket{-}
\ket{+}\Big), \eqno{(21)}$$
where $\ket{\buildrel + \over -}$ are
the spin-up and the spin-down state
vectors respectively.

It is straightforward to see that the
three nonsinglet Bell states
$\ket{\psi_s}$, $s=1,2,3$, when written
in the form (20), are given by
$t_s=-1$, and the other two components
of $\vec t$ equal to $+1$. The singlet
state $\ket{\psi_0}$ is in the form
(20) determined by all three components
of $\vec t$ being equal to $-1$.

It is also easy to see that for all
mixtures one has
$$ -1\leq t_i\leq +1\qquad i=1,2,3.$$
This is a necessary, but not a
sufficient condition for $T$ being a
state. In other words, the tetrahedron
is embedded in a cube, in which there
are also nonphysical $\vec t$. In view
of the LHS of (20), we call $T$ that
belong to the tetrahedron: {\it
generating MDS states}.

What we want to find out is: {\it Which
of the MDS  states have nontrivial
twins?} For those that do have, we want
to find {\it the set of all nontrivial
pairs of twins}.

It is sufficient to find the generating
MDS  states $T$ with nontrivial twins,
because the validity of
$$ A_1T=A_2T $$
obviously implies
$$ \Big(U_1A_1U_1^{\dagger}\Big)
\Big(U_1U_2TU_1^{\dagger}
U_2^{\dagger}\Big)=\Big(U_2A_2U_2^{\dagger}
\Big)\Big(U_1U_2TU_1^{\dagger}
U_2^{\dagger}\Big), $$
i. e., if the generating MDS  states
have nontrivial twins, then also the
generated MDS  states do have
nontrivial twins, and they are
immediately obtained.

As far as the pure generating MDS
states (the Bell states) are concerned,
the first-particle reduced statistical
operator $\rho_1$ is equal to
$(1/2)I_1$, {\it all} nontrivial
Hermitian operators $A_1$ commute with
it, hence \cite{FV76}, they are twins.
To evaluate the corresponding twin
$A_2$, one has to read off the
antilinear correlation operator $U_a$
from (21) having in mind (11), and then
utilize (13). For the best known Bell
state, the singlet state
$\ket{\psi_0}$, e. g., $U_a$ takes
$\ket{+}$ into $\ket{-}$, and $\ket{-}$
into $(-\ket{+})$ (cf (21)). If
$$ A_1=\alpha_{++}\ket{+}\bra{+}\enskip
+\enskip
\alpha_{--}\ket{-}\bra{-}\enskip
+\enskip \alpha_{+-}
\ket{+}\bra{-}\enskip +\enskip
(\alpha_{+-})^*\ket{-} \bra{+},$$
$$\alpha_{++},\alpha_{--}\in \mbox{\bf
R},\quad \alpha_{+-}\in \mbox{\bf C},$$
then the twin $A_2$ has the form:
$$ A_2=\alpha_{--}\ket{+}\bra{+}\enskip
+\enskip
\alpha_{++}\ket{-}\bra{-}\enskip
-\enskip \alpha_{+-}
\ket{+}\bra{-}\enskip -\enskip
(\alpha_{+-})^*\ket{-} \bra{+}.$$

Now we turn to the {\it mixtures of
Bell states} in our search for
nontrivial twins.\\

\section{Mixtures of
Bell states}

 Viewing statistical operators as super
vectors, and utilizing (redundantly,
but for the sake of better overview)
the ket notation for super state
vectors (i. e., Hilbert-Schmidt
operators as normalized super vectors),
one can rewrite the generating vectors
$T$ given by (20) as a biorthogonal
expansion with positive expansion
coefficients:
$$ \ket{T\|T\|^{-1}}_{12}=
(1+\sum_{i=1}^3t_i^2)^{-1/2}
\Big(\ket{(1/2)^{1/2}I}_1\otimes
\ket{(1/2)^{1/2}I}_2+$$
$$\sum_{i=1}^3|t_i|
\ket{(1/2)^{1/2}\sigma_i}_1 \otimes
\ket{sg(t_i)(1/2)^{1/2}\sigma_i}_2\Big)
\eqno{(22)}$$
("$sg$" denotes the sign), i. e., as a
(super state vector) {\it Hermitian
Schmidt decomposition}.

One can read off (22) the following
canonical entities of the super state
vector $ \ket{T\|T\|^{-1}}_{12}$ (cf
(10a), (10b) and (11)):

The first-subsystem reduced statistical
super operator $\hat \rho_1$ has the
characteristic super state vectors
$\{\ket{(1/2)^{1/2}I}_1,\enskip
\ket{(1/2)^{1/2}\sigma_i}_1:\enskip
i=1,2,3\}$; the second subsystem
reduced statistical super operator
$\hat \rho_2$ has the characteristic
state vectors
$\{\ket{(1/2)^{1/2}I}_2,\enskip
\ket{sg(t_i)(1/2)^{1/2}
\sigma_i}_2:\enskip i=1,2,3\}$; and the
common spectrum of $\hat \rho_1$ and
$\hat \rho_2$ is $\{R_0\equiv
(1+\sum_{i=1}^3 t_i^2)^{-1},\enskip
R_i\equiv R_0t_i^2:\enskip i=1,2,3\}$.
Finally, the antiunitary correlation
super operator $\hat U_a$ maps the
enumerated characteristic state vectors
of $\hat \rho_1$ into the
correspondingly ordered ones of $\hat
\rho_2$.

\section{Nontrivial MDS  twins}
Every super operator $\hat A_1$ that
commutes with $\hat \rho_1$, i. e., for
which every characteristic subspace of
the latter is invariant (and no other
super operator ), has a twin super
operator $\hat A_2$ \cite{FV76}. But we
are interested only in those pairs
$(\hat A_1,\hat A_2)$ in which {\it
both} super operators are, what may be
called, {\it multiplicative} ones, i.
e., which have the form
$$ \hat A_1\rho_{12}=A_1\rho_{12},\quad
\hat A_2\rho_{12}=A_2\rho_{12}, $$
where $A_p$, $p=1,2$, are ordinary
(subsystem) operators. It is easy to
see that a multiplicative super
operator is {\it Hermitian} (in the
Hilbert-Schmidt space of supervectors)
if so is the ordinary operator (in the
usual sense) that determines it.

The basic result of the expounded
illustration is given in the following
two theorems:\\

\noindent {\it Theorem 5.} Mixed
generating MDS states have nontrivial
twins if and only if they are mixtures
of two Bell states (binary mixtures).\\

\noindent {\it Theorem 6.} A) Let us
take a binary mixture of two Bell
states both distinct from the singlet
one, and let $T_i\equiv
\ket{\psi_i}\bra{\psi_i}$ (cf (21)) be
the nonsinglet Bell state that does not
participate in the mixture. Then the
nontrivial twins are:
$$ A_1\equiv \alpha I_1 +\beta
\sigma_i^{(1)},\qquad A_2\equiv \alpha
I_2 +\beta \sigma_i^{(2)},\qquad \alpha
,\beta \in \mbox{\bf R},\enskip \beta
\not= 0, \eqno{(23)}$$
where the suffix on $\sigma_i$ refers
to the corresponding tensor factor
space.

 B) In case of a binary mixture
of the singlet state with another Bell
state, say $T_i\equiv
\ket{\psi_i}\bra{\psi_i}$ (cf (21)),
the twins are:
$$ A_1\equiv \alpha I_1 +\beta
\sigma_i^{(1)},\qquad A_2\equiv \alpha
I_2-\beta \sigma_i^{(2)},\qquad \alpha
,\beta \in \mbox{\bf R},\enskip \beta
\not= 0. \eqno{(24)}$$

Proof of the two theorems and of some
subsidiary results is given in the
Appendix. The proof of Theorem 6 that
is first given in the Appendix is only
of {\it methodological significance}:
it illustrates a method how to evaluate
nontrivial twins. In our case of binary
mixtures $T^{(2)}$, another method
gives a simpler evaluation. It is given
at the end of the Appendix.

It is known that any Bell state can be
converted into any other one by local
unitary transformation \cite{Ben},
\cite{Ben'}. Hence, for proving the
existence of nontrivial twins it would
have sufficed to take mixtures of one
pair of Bell states.  Theorem 6 is,
nevertheless, more elaborate because
the explicit form of the twins depends
on which Bell states are involved.

It is known that all binary mixtures of
Bell states are nonseparable except
those with equal weights. The latter,
as easily seen, are examples of Theorem
4, e. g.,  as one can easily ascertain
making use of (21), one has:
$$ (1/2)\Big((\ket{+}\bra{+}\otimes
\ket{+}\bra{+})\enskip +\enskip
(\ket{-}\bra{-}\otimes
\ket{-}\bra{-})\Big)=$$
$$ (1/2)\Big(\ket{\psi_1}\bra{\psi_1}
+\ket{\psi_2}\bra{\psi_2}\Big).
\eqno{(25)}$$

The nonseparable binary Bell state
mixtures are distillable even in the
single copy case \cite{Hor97}.
Unfortunately, there is no simple
relation between the (investigated)
existence of nontrivial twins and
distillability as seen from the fact
that also rank four mixtures are
distillable (if and only if one of the
weights is larger than $1/2$), and they
do not have nontrivial twins.

In conclusion, I would like to point
out that the entangled pure state case
\cite{FV76}, \cite{VF84} is a well
explored illustration for the fact that
nontrivial twins can exist on account
of {\it entanglement}. The nonseparable
binary Bell state mixtures are another
simple illustration for this fact. One
should have in mind that, as it was
seen in Lemma 3, uncorrelated bipartite
states do not have nontrivial twins.
Separable states can have nontrivial
twins if and only if biorthogonal
grouping of the terms is possible (cf
Theorem 4). Unfortunately, for the time
being, we do not have a necessary and
sufficient condition for the existence
of nontrivial twins on account of
entanglement (except in the pure state
case), let alone a way of generating
all of them for a given
composite-system mixed state (except in
the pure state case).

At last, but not at least, a relation
between the reported twin investigation
(\cite{FV76}, \cite{VF84}, and
\cite{HD00} besides this article) and
the mainstream research on entanglement
(take the cited Bennett et al.
articles, the article of Vedral et al.,
and the cited Horodecki family articles
as examples) is still lacking. But I
believe that there exists a connection.
Further research will, hopefully,
uncover it.

\section{Appendix}
Since we are going to prove the
theorems making use of (22), first we
must be able to recognize the binary
mixtures $T^{(2)}$ on the Horodecki
tetrahedron.\\

\noindent {\it Proposition A.1.} One
has a binary mixture $T^{(2)}$ if and
only if precisely one of the three
$|t_i|$ values in (22) equals $1$.

A) If $t_i=+1,\enskip |t_{i+1}|,
|t_{i+2}|<1$ (where the three values
$\{1,2,3\}$ of $i$ are meant
cyclically), then the mixture is of two
Bell states both distinct from the
singlet state. If $T_i$ is the
nonsinglet Bell state that does not
participate in the mixture, one has
$t_{i+2}= -t_{i+1}$. Finally, the
binary mixture $T^{(2)}$ in question is
$$
T^{(2)}=\Big[(1-t_{i+1})/2\Big]T_{i+1}
+\Big[(1-t_{i+2})/2\Big]T_{i+2}.\eqno{(A1)}$$

B) If $t_i=-1,\enskip |t_{i+1}|,
|t_{i+2}|<1$ (in the cyclic sense),
then one deals with a mixture of two
states: the singlet state and another
Bell state $T_i$. One has $t_{i+1}
=t_{i+2}$, and the binary mixture
$T^{(2)}$ in question is
$$ T^{(2)}=\Big[(1+t_{i+1})/2\Big]T_i+
\Big[
(1-t_{i+1})/2\Big]T_0.\eqno{(A2)}$$

Both in the cases (A) and (B),
$t_{i+1}$ can be any number in the
interval $-1\leq t_{i+1}\leq +1$;
equivalently, one can have any point on
the corresponding border of the
Horodecki tetrahedron (the vertices
excluded).\\

For proof a few subsidiary results are
required.\\

\noindent {\it Lemma A.1.} If among the
four numbers $\{1,|t_i|:i=1,2,3\}$
appearing in the form (22) of the
generating MDS state $T$ there is one
distinct from the rest, then $T$ has no
nontrivial twins.\\

\noindent {\it Proof.} As clearly
follows from the above stated spectrum
of $\hat \rho_1$, the mentioned "one
number distinct from the rest"
corresponds to a nondegenerate
characteristic value. Assuming that
$A_1$ is a twin, it is a multiplicative
superoperator reducing in each
characteristic subspace of $\hat
\rho_1$. (This is equivalent to
commutation with $\hat \rho_1$.)

    a) Let us take the case when
$|t_i|<1,\quad i=1,2,3$. Then the first
characteristic value of $\hat \rho_1$
is nondegenerate, and the corresponding
characteristic super state vector has
to be invariant (up to a constant):
$$ A_1(1/2)^{1/2}I_1=\alpha
(1/2)^{1/2}I_1,$$
i. e., $A_1=\alpha$, and the twin is
trivial.

b) Let $|t_i|$ for some value of $i$ be
distinct from the other three numbers.
Then the corresponding characteristic
super state vector
$\sigma_i(1/2)^{1/2}$ must be invariant
(up to a constant):
$$ A_1\sigma_i(1/2)^{1/2}=\alpha
\sigma_i(1/2)^{1/2}, $$
which, upon multiplication with
$\sigma_i$ from the right, implies
$A_1=\alpha$ again.\hfill $\Box$ \\

\noindent {\it Corollary A.1.} If a
generating MDS state $T$ has nontrivial
twins, then for at least one value of
$i$: $|t_i|=1$.\\

\noindent {\it Proof} is obvious from
Lemma A.1.\hfill $\Box$ \\

\noindent {\it Lemma A.2.} Expressing a
generating MDS  state $T$ written in
the form (22) in terms of the
statistical weights with respect to the
Bell states $\{T_k\equiv
\ket{\psi_k}\bra{\psi_k}:k=0,1,\dots
,3\}$ (cf (6)), one has:
$$ T=\sum_{k=0}^3w_kT_k= (1/4)
\Big[I\otimes
I+(-w_1+w_2+w_3-w_0)\sigma_1\otimes
\sigma_1+(w_1-w_2+w_3-w_0)\sigma_2\otimes
\sigma_2+$$
$$(w_1+w_2-w_3-w_0)\sigma_3\otimes
\sigma_3\Big],\eqno{(A3)}$$
where
$$ \forall k:\quad w_k\in [0,1],\quad
k=0,1,2,3;\quad \sum_{k=0}^3w_k=1.$$

\noindent {\it Proof} is
straightforward substituting the Bell
states in (22) (cf (21) and beneath
it).\hfill $\Box$ \\

\noindent {\it Lemma A.3.} If  one has
$|t_i|=1,\enskip i=1,2,3$, for a
generating MDS state $T$ in the form
(22), then it is a Bell state.\\

\noindent {\it Proof.} Each $t_i$ has
two sign possibilities; altogether
there are $2^3=8$ possibilities. A
straightforward analysis of each of
these, taking into account Lemma A.2
and $\sum_{k=0}^3w_k=1$, shows that $4$
possibilities do not give states. These
are: $\{sg(t_i)=+:i=1,2,3\},\enskip
\{+--\},\enskip \{-+-\}$, and
$\{--+\}$. The remaining four sign
possibilities give the four Bell
states:
$$ \{-++\}:\enskip T_1;\quad
\{+-+\}:\enskip T_2;\quad
\{++-\}:\enskip T_3;\quad
\{---\}:\enskip T_0.$$
\hfill  $\Box$

\noindent {\it Proof of claim (A) in
Proposition A.1.}  Since it is clear
from (A3) that the $t_i$ as functions
of $w_k$ are symmetric (in the sense of
the cycle $\{1,2,3\}$), it is
sufficient to take $i=1$. Then
$$ -w_1+w_2+w_3-w_0=1,\quad \mbox{and}
\quad \sum_{k=0}^3w_k=1. $$
This gives $w_2+w_3=1,\enskip
w_1=w_0=0$, and $t_2= w_3-w_2=-t_3$.
Hence, $w_2= (1-t_2)/2$ and
$w_3=(1+t_2)/2$ as claimed. Since $0<
w_1,w_0<1$, the claimed intervals for
$t_2$ and $t_3$ follow.\hfill $\Box$ \\

\noindent {\it Proof of claim (B) of
the Proposition.} It runs in full
analogy with the proof for case
(A).\hfill $\Box$ \\

\noindent {\it Proof of the main claim
of the Proposition.} It is easy to see
that the proofs of claims (A) and (B)
of the Proposition go through also for
the case when $|t_{i+1}|$ or
$|t_{i+2}|$ equals one. Hence, one
cannot have $|t_i|=1$ for precisely two
values of $i$. If it is so for one
value, then either it is so for all
three values, and one has a pure Bell
state, or it is so for precisely one
value of $i$, then we have a binary
mixture.\hfill $\Box$
\\

\noindent {\it Proof of Theorem 6.} We
now assume that for one value of $i$,
$|t_i|=1$, and that the other two
components of $\vec t$ in (22) are by
modulus less than one. Then it is
sufficient and necessary for an
observable $A_1$ that defines a
superoperator $\hat A_1$ by
multiplication (we write this as $\hat
A_1\equiv (A_1\bullet )$) to have a
superoperator twin $A_2$ (that is not
necessarily multiplicative as $\hat
A_1$) that it reduces in the
two-dimensional supervector subspace
spanned by $I_1$ and $\sigma_i^{(1)}$.
If we write $A_1=\alpha I_1
+\sum_{j=1}^3\beta_j\sigma_j^{(1)}$
($\alpha ,\beta_j\in${\bf R}), and
multiply with this from the left
$\sigma_i^{(1)}$, it turns out that the
condition amounts to $\beta_j=0,\enskip
j\not= i$. The symmetrical argument
gives the symmetrical result. Thus the
multiplicative superoperators defined
by $A_1$ and, separately, by $A_2$ do
have superoperator twins if and only if
they are of the form
$$A_1=\alpha I_1+\beta
\sigma_i^{(1)},\qquad A_2=\gamma I_2
+\delta \sigma_i^{(2)},\eqno{(A4)}$$
where $\alpha ,\beta ,\gamma ,\delta
\in \mbox{\bf R}$.

 The mentioned operators are
twins of each other if and only if
$$ (A_2\bullet )=\hat U_a(A_1\bullet )
\hat U_a^{-1}.\eqno{(A5)}$$

Now we find out the necessary and
sufficient conditions when (A5) is
valid for the operators given by (A4).
Since both sides of (A5) are linear
operators, we apply them to the basis
of supervectors
$\{I_2,\sigma_i^{(2)}:i=1,2,3\}$:
$$ (A_2\bullet )I_2=\gamma I_2+\delta
\sigma_i^{(2)};$$
$$ \Big(\hat U_a(A_1\bullet )\hat
U_a^{-1}\Big)I_2=\hat U_a(\alpha I_1+
\beta \sigma_i^{(1)})=\alpha I_2+
sg(t_i)\beta \sigma_i^{(2)}.$$

Thus, we obtain the condition
$$ \gamma =\alpha ,\quad \delta=
sg(t_i)\beta .$$

Utilizing the well known relation
$$ \sigma_i\sigma_j=\delta_{ij}I+
\sum_{m=1}^3i\epsilon_{ijm}\sigma_m,$$
we, further, have
$$ (A_2\bullet )\sigma_j^{(2)}=(\gamma
I_2+ \delta \sigma_i^{(2)})
\sigma_j^{(2)}=\gamma \sigma_j^{(2)}+
\delta (\delta_{ij}I_2+\sum_mi
\epsilon_{ijm}\sigma_m^{(2)});$$
$$ \Big(\hat U_a(A_1\bullet )\hat
U_a^{-1}\Big)\sigma_j^{(2)}=sg(t_j)\hat
U_a(\alpha I_1+\beta \sigma_i^{(1)})
\sigma_j^{(1)}=$$
$$ sg(t_j)\hat U_a\Big( \alpha
\sigma_j^{(1)}+\beta
(\delta_{ij}I_1+\sum_mi\epsilon_{ijm}
\sigma_m^{(1)})\Big)=$$
$$sg(t_j)\Big(\alpha sg(t_j)
\sigma_j^{(2)}+\beta (\delta_{ij}I_2-
\sum_mi\epsilon_{ijm}sg(t_m)
\sigma_m^{(2)})\Big).$$
For $i=j$ we obtain the condition
$\gamma =\alpha$, and $\delta
=sg(t_i)\beta$, and, for $j\not= i$, in
addition: $\delta =-sg(t_j)
sg(t_m)\beta$. Since $i\not= m\not= j$,
we know from the Proposition that,
irrespective of $sg(t_i)$, one has
$-sg(t_j)sg(t_m)=sg(t_i)$. Hence, we
actually obtain the condition expressed
by (23) and (24).\hfill $\Box$

The claim in Theorem 5 that binary
mixtures $T^{(2)}$ do have nontrivial
twins is an immediate consequence of
Theorem 6.\\

The mentioned second, simpler, proof
goes as follows:  According to Lemma 1,
a pair of opposite-subsystem
observables $(A_1,A_2)$ are twins for a
composite-system mixture if and only if
they are simultaneously twins for each
of the pure term states.

Utilizing (13), it is straightforward
to evaluate the twins in the operator
basis consisting of the four
supervectors
$\ket{\stackrel{+}{-}}\bra{\stackrel{+}{-}}$.
But for comparison with the results
(23) and (24) obtained by the Hermitian
Schmidt decomposition method, we do
this in a little bit more difficult way
using the form (22) for the Bell states
(see their description beneath (22)).

We can read off the antiunitary
correlation superoperator $\hat U_a$
from the mentioned form (22) of the
Bell state. As it was stated before,
every first-subsystem observable
$A_1\equiv \alpha I_1
+\sum_{i=1}^3\beta_i\sigma_i^{(1)},\enskip
(\alpha ,\beta_i\in \mbox{\bf
R},\enskip i=1,2,3)$, is a twin. The
corresponding second-subsystem twins
for the Bell states are:
$$ T_1:\quad A_2\equiv \alpha I_2
-\beta_1\sigma_1^{(2)}+\beta_2
\sigma_2^{(2)}
+\beta_3\sigma_3^{(2)};$$
$$ T_2:\quad A_2\equiv \alpha I_2
+\beta_1\sigma_1^{(2)}-\beta_2
\sigma_2^{(2)} +
\beta_3\sigma_3^{(2)};$$
$$ T_3:\quad A_2\equiv \alpha
I_2+\beta_1 \sigma_1^{(2)}+
\beta_2\sigma_2^{(2)}- \beta_3
\sigma_3^{(2)};$$
$$ T_0:\quad A_2\equiv \alpha
I_2-\beta_1 \sigma_1^{(2)}-
\beta_2\sigma_2^{(2)}- \beta_3
\sigma_3^{(2)}.$$

Now, in view of the position of the
minus sign in $A_2$, evidently,
utilizing $m\not= i\not= j\not=
m\enskip i,j,m\in \{1,2,3\}$, and
$0<w<1$, the simultaneous twins are:
$$ wT_j+(1-w)T_m:\quad A_2\equiv \alpha
+\beta_i\sigma_i;$$
$$ wT_0+(1-w)T_i:\quad A_2\equiv \alpha
-\beta_i\sigma_i;$$
and $A_2$ is,  of course, the twin of
$A_1\equiv \alpha +\beta_i\sigma_i$.

In this way proof of (23) and (24) is
obtained.\\

\noindent {\bf Acknowledgements}\\

\noindent The author is grateful to
Prof. Anton Zeilinger for his
invitation. Thanks are due also for the
financial support and hospitality of
the "Erwin Schr\"{o}dinger" Institute
in Vienna, where part of this work was
done. The author is indebted to
\v{C}aslav Brukner, who was of help by
asking the right question.



\end{document}